\documentclass{emulateapj}
\def\hbeta{H$\beta$}

\def\rlow{ \ifmmode r_{54} \else $r_{54}$\fi}
\def\rmed{ \ifmmode r_{42} \else $r_{42}$\fi}
\def\rhigh{\ifmmode r_{21} \else $r_{21}$\fi}
\def\rall{\ifmmode r_{51}  \else $r_{51}$\fi}

\def\msun{\ifmmode {\rm M_\odot} \else M$_\odot$\fi}
\def\msunyr{\ifmmode {\rm M_\odot~yr^{-1}}\else${\rm M_\odot~yr^{-1}}$\fi}

\def\lam{\ifmmode {\lambda} \else {$\lambda$}\fi}
\def\cosi{\ifmmode {\cos\,i} \else $\cos\,i$\fi}

\def\teff{\ifmmode {T_{\rm eff}} \else $T_{\rm eff}$\fi}
\def\tmax{\ifmmode {T_{\rm max}} \else $T_{\rm max}$\fi}
\def\mbh{\ifmmode {M_{\rm BH}} \else $M_{\rm BH}$\fi}
\def\mdot{\ifmmode {\dot M} \else $\dot M$\fi}
\def\mdoto{\ifmmode {\dot{M}_0} \else  $\dot{M}_0$\fi}
\def\lamLlam{\ifmmode \lambda L_{\lambda}(5100) \else {$\lambda
    L_{\lambda}(5100)$} \fi}

\def\llam{\ifmmode {L_\lambda} \else  $L_\lambda$ \fi}
\def\flam{\ifmmode {F_\lambda} \else  $F_\lambda$ \fi}
\def\sigstar{\ifmmode \sigma_* \else $\sigma_*$\fi}
\def\hbeta{\ifmmode {\rm H}\beta \else H$\beta$\fi}
\def\civ{\ifmmode {\rm C{\sc iv}} \else C~{\sc iv}\fi}

\def\ergps{\ifmmode \rm erg~s^{-1} \else $\rm erg~s^{-1}$ \fi}
\def\kmps{\ifmmode \rm km~s^{-1}\else $\rm km~s^{-1}$\fi}
\def\kms{km~s$^{-1}$}

\def\lbol{\ifmmode {L_\mathrm {bol}} \else $L_\mathrm {bol}$\fi}
\def\led{\ifmmode {L_\mathrm {Ed}} \else $L_\mathrm {Ed}$\fi}

\def\mgii{\ifmmode {\rm Mg{\sc ii}} \else Mg~{\sc ii}\fi}

\def\gsim{\lower 2pt \hbox{$\, \buildrel {\scriptstyle >}\over
{\scriptstyle \sim}\,$}}
\def\lsim{\lower 2pt \hbox{$\, \buildrel {\scriptstyle <}\over
{\scriptstyle \sim}\,$}}

\newcommand{\oiii}{{\sc [O~iii]}}

\newcommand{\feii}{Fe~{\sc ii}}

\slugcomment{Submitted to The Astrophysical Journal}

\shorttitle{Accretion Disk Temperature -- Continuum Color relationship}
\shortauthors{Bonning et al.}

\begin{document}

\title{Accretion Disk Temperatures and Continuum Colors in QSOs}

\author{E.~W. Bonning}
\affil{Laboratoire de l'Univers et de ses Th\'{e}ories, Observatoire
  de Paris, F-92195 Meudon Cedex, France} 

\and
\author{L. Cheng,  G.~A. Shields, S. Salviander, K. Gebhardt}
\affil{Department of Astronomy, University of Texas, Austin, TX 78712}

\begin{abstract}

Accretion disks around supermassive black holes are widely believed to 
be the dominant source of the optical-ultraviolet continuum in many
classes of active  galactic nuclei (AGN).   We study here the  
relationship between the continuum colors of AGN and the
characteristic  accretion disk temperature ($T_{\rm max}$). Based on NLTE
models of  accretion disks in AGN computed as described
by Hubeny et al. (2000),  we 
find that continuum intensity ratios for several pairs
of wavelengths between 1350 and 5100~\AA\ 
should show a
trend of bluer colors for  higher $T_{\rm max}$, notwithstanding
random disk inclinations. We compare this theoretical expectation
with observed colors of QSOs in the  Sloan Digital Sky Survey (SDSS),
deriving black hole mass and thence $T_{\rm max}$  from the width of the
\mgii\ broad emission line. The observed colors generally do not show
the expected trend and in some cases show a reverse trend of redder
colors with increasing \tmax.
The cause of this discrepancy does not appear to be dust reddening or
galaxy contamination but may relate to the accretion rate,  as the
offset objects are accreting above $\sim$ 30 \% of the Eddington limit.
The derived disk temperature depends primarily on line width, with
little or no dependence on  luminosity.

\end{abstract}


\section{Introduction}
Accretion onto supermassive black holes is widely accepted as the
energy source of AGN.  Much of the optical and ultraviolet luminosity,
specifically the so-called ``Big Blue Bump''  may be thermal
emission from an accretion disk (Shields 1978; Malkan 1983; for a
review of attempts to relate disk models to QSO spectra see Koratkar \&
Blaes 1999).   In the simplest picture, radiation from each radius in
the disk is powered by energy produced locally by viscous dissipation
within the vertical thickness of the disk.  Models achieve some
success in fitting the spectral energy distribution (SED) of
individual QSOs (e.g., Sun \& Malkan 1989; Blaes et al. 2001; and
references therein). Simple disk models nevertheless have shortcomings
(e.g., Antonucci 1999).   
Time variations of the optical, ultraviolet, and X-ray continuum of
some AGN show patterns suggestive of reprocessing in an 
externally illuminated atmosphere (Collin 1991).
Polarization of the optical continuum of QSOs is weaker in degree and
opposite in sign  from that expected from a smooth electron scattering 
atmosphere (see Coleman \& Shields 1990, and references therein).  The
soft X-ray continuum cannot be explained by ordinary thermal emission
from the disk atmosphere, and   
requires another ingredient such as a Compton-scattering corona
(Wilkes \& Elvis 1987; Comastri et al. 1992; Blaes et al. 2001). 
On the other hand, Kishimoto et al. (2005) argue from
spectropolarimetry that the infrared continuum of AGN accretion disks
has a slope similar to that expected from simple disk models.

The mixed success of the disk model makes it important to explore
further tests of the basic agreement between the model and observations.
One simple test involves the temperature of the disk and the expected
color of the continuum.  Wien's law suggests
that hotter disks should in general have bluer continua.  This is born out by
models, as we discuss below.  The disk effective temperature varies
with radius in a self-similar way, and the temperature of a given disk can be
parameterized by the maximum value of the local effective temperature,
$T_{\rm max} \propto L_{\rm bol}^{1/4}  \mbh^{-1/2}$, that occurs at a radius
slightly outside the inner disk boundary (see review by Novikov \&
Thorne 1973).  \tmax\ varies with \mbh\ and the
accretion rate \mdot\ as $\tmax^4 \propto \mdot/\mbh^2$, because the
bolometric luminosity ( L$_{\rm bol}$) is proportional to \mdot and the
disk radius scales with the gravitational radius $R_g = GM/c^2$.  One
consequence is that, for objects shining at a given fraction of the
Eddington limit ($\led \propto \mbh$), the temperature decreases with
increasing luminosity or mass. On this basis, one might expect
luminous QSOs to have softer (redder) continuum slopes than lower
luminosity AGN. This expectation is not consistently fulfilled by
observed SEDs of AGN. There is a trend of increasing soft X-ray excess
with narrower \hbeta\ and higher Eddington ratio (e.g., Laor et
al. 1997); these quantities correlate in practice with higher \tmax.  
The ``Baldwin effect'', which involves decreasing emission-line
equivalent width for higher  luminosity, is also consistent (Wandel
1999; Dietrich et al. 2002). More luminous objects should have cooler
disks and softer ionizing continua. Scott et al. (2004)
find harder EUV slopes for higher \mbh\ and luminosity, although
curiously the trend is largely lost in scatter in their plot of slope
versus \tmax. Blaes (2003), looking at continuum slope as a function
of  \mbh\ and $L/L_{\rm Ed}$ for a small set of AGN, found a large degree
of scatter but  an overall trend apparently opposite to theoretical
expectation (also noted by Cheng, Shields, \& Gebhardt 2004).

\tmax\ may be estimated for individual objects with knowledge of \mbh\
and \mdot. The accretion rate can be estimated from the bolometric
luminosity, and \mbh\ can be estimated with the aid of recently
established methods utilizing the  widths of the broad emission lines
(Kaspi et al. 2000, 2005; Bentz et al. 2006).  In this paper, we use
spectra from the SDSS DR3 \footnote{The SDSS website is
http://www.sdss.org.} to examine the continuum slope of QSOs as a
function of \tmax.    In Section
\ref{sec:work}, we describe the theoretical models we use and compare
them to observed QSOs.  Results are discussed in Section
\ref{sec:discussion}.

\section{Modeling and Data Sampling}
\label{sec:work}
\subsection{Accretion Disk Properties}
\label{sec:thorne}
The Eddington limit for a black hole (BH) with mass $\mbh$ is given by
\begin{equation}
 L_\mathrm{Ed} = \frac{4\pi cG\mbh}{\kappa_{e}}=(10^{46.10} \ergps)M_8,
\end{equation}
where $\kappa_{e}$ is the electron scattering opacity per unit mass
and M$_8 = \mbh / (10^8~\mathrm M_{\odot})$.
The disk luminosity production through accretion onto the black hole is
\begin{equation}
 L = \epsilon \dot{M} c^{2} = (10^{45.76}{\rm erg~s^{-1}})\epsilon_{-1}
 \dot{M}_{\mathrm 0} .
\end{equation}
where $\dot{M}_{\mathrm 0}$ is the accretion rate in \msunyr, and
$\epsilon$ = 10$\epsilon_{-1}$ is the  efficiency. For a Schwarzschild
(non-rotating) black hole, $\epsilon$ = 0.057, and  for a rapidly
rotating Kerr hole with angular momentum parameter $a_{\ast}$ = 0.998,
$\epsilon$ = 0.31 (Novikov \& Thorne 1973).  The effective temperature
at radius $r$ is given by the local radiative flux leaving the
atmosphere, $\teff = (F/\sigma)^{1/4}$.   For $a_{\ast}$ = 0.998
(Thorne 1974),
\begin{equation}
 \tmax=(10^{5.56}~{\rm K})M_{8}^{-1/4}(\lbol/\led)^{1/4},
\label{eq:tmax1}
\end{equation}
or alternatively,
  $\tmax=(10^{5.54}~{\rm K})M_{8}^{-1/2}\, L_{46}^{1/4}$, where
  $L_{46} \equiv \lbol/(10^{46}~\ergps)$. 
For a Schwarzschild hole, $T_\mathrm {max}$ is  cooler by a factor of
$10^{0.46}$ for a given \mbh\ and \lbol\ (e.g., Shields 1989). 

In traditional treatments, the local vertical structure is determined
by hydrostatic equilibrium and radiative and convective transport of
heat produced locally by viscous  stresses.  The viscosity also
determines the radial diffusion of matter and the surface density of
matter in the disk as a function of radius.  The locally emitted
spectrum is determined by \teff\ and the effective 
gravity in the atmosphere.  The locally emitted spectrum shows strong
features such as the Lyman edges of hydrogen and helium, but in the observed
spectrum these features are  broadened by relativistic
effects.

\subsection{Deriving \mbh\ and \tmax }
\label{sec:tmax}
Derivation of \mbh\ from the width of the \hbeta\ or \mgii\ broad
emission lines has become widely accepted in recent years.  The FWHM
of the broad lines is taken to give the circular velocity of the
broad-line emitting material (with some geometrical correction
factor).  The radius of the broad-line region (BLR), derived from echo
mapping studies, increases as a function of the continuum luminosity,
$R \propto L^{\Gamma}$ with $\Gamma = 0.5 - 0.7$ (Wandel et al. 1999;
Kaspi et al. 2000, 2005).   Simple considerations involving
photoionization physics suggest $\Gamma = 0.5$ (Shields et al. 2003).
Bentz et al. (2006) find $\Gamma = 0.52\pm0.04$ after correcting the
sample of Kaspi et al. (2005) for host galaxy contamination.  Such
studies typically use the \hbeta\ broad line, but McLure \& Jarvis
(2002) find that the \mgii~$\lambda2800$ line is a valid alternative
useful at higher redshifts. Here we use the expression for \mbh
\begin{equation}
\label{eq:mbh}
\mbh = (10^{7.69}~\msun)v_{3000}^2 L_{44}^{\Gamma},
\end{equation}
where $v_{3000} \equiv \rm{FWHM}$/3000~\kmps, $\Gamma=0.5$,
$ L_{44}~\equiv~\lamLlam/(10^{44}~\ergps)$, and
the coefficient is taken from Shields et al. (2003).  We use the FWHM of the
\mgii\ line in order to get maximum coverage in redshift for a single
broad emission line.  

The derived value of \tmax\ depends almost entirely on the the broad
line FWHM for a given object.  The bolometric luminosity can be
estimated as $\lbol = f_L \times \lamLlam$, following Kaspi et
al. (2000). Using  Eq.~\ref{eq:mbh} and this expression for \lbol,
Eq.~\ref{eq:tmax1} becomes
\begin{equation}
\label{eq:tmax2}
\tmax = (10^{5.43}~K)~v_{3000}^{-1}~L_{44}^{-(\Gamma -
0.5)/2}~(f_L/9)^{-1/4},
\end{equation}
Note that for $\Gamma = 0.5$, as used here, the derived disk temperature is
inversely proportional to FWHM with no dependence on $L$,
save the bolometric correction factor $f_L=9$ (Kaspi et al. 2000).
Shang et al. (2005), in an analysis of the SEDs of 17 well observed
QSOs, also adopted $f_L = 9$, although their broad band SEDs typically indicate
a larger value, on average $f_L \approx 13$. More recently,
  Richards et al. (2006) find an average bolometric correction from
  5100~\AA~ of $\sim$10 from composite SEDs of 259 QSOs.  We
discuss the sensitivity of our results to the value of $f_L$ in
Sec.~\ref{sec:param} below. 
\subsection{The Color-$T_\mathrm {max}$ Relation from the Non-LTE
  Models}
\label{sec:models} 
One might expect that cooler disks will in general have
redder colors. This certainly  would be true (apart from inclination
effects) for non-relativistic disks emitting locally as black bodies.
However, disks in AGN are expected to depart from black body emission
locally, and the energy distribution received by a 
particular observer is strongly affected by relativistic effects.  The
observer's angle $i$ with respect to the disk axis is also important,
because the radiation from the innermost, hottest part of the disk
will be relativistically beamed near the equatorial plane, especially for 
rapidly rotating holes (Cunningham 1975; Laor \& Netzer
1989). Therefore, models are needed to establish the expected
statistical relationship between \tmax\ and color.  In
recent years, I. Hubeny and his colleagues have computed NLTE disk
models for AGN and made comparisons with observation 
(Hubeny et al. 2000).  Hubeny et al. have kindly made available
the computer program {\sc agnspec} which gives the observed energy 
distribution ($4\pi dL_{\nu}/d\Omega$) emitted in the observer's
direction for model disks as a function of \mbh, \mdot, $a_*$, 
inclination, and the Shakura \& Sunyaev (1973) viscosity parameter $\alpha$.
The vertical structure and local emission is evaluated at each radius
in the disk, and the spectrum viewed at infinity is calculated via the
general relativistic transfer function.  Using {\sc agnspec}, we computed a
set of models with values of \mbh\ and \mdot\ representative of our
SDSS QSO sample (discussed below). We select the range of
  \mdot to give Eddington ratios consistent with those in our observed
sample (see Fig.~\ref{fig:lled} and discussion below. We fix $\alpha
= 0.10$ and, 
except as noted below, $a_* = 0.998$.  Hubeny et al. (2000) show
that models with $\alpha = 0.1$ and 0.01 differ negligibly at the
wavelengths of interest here. To assess the effect of disk inclination
on the observed colors, we assign a random inclination (uniform
distribution in \cosi) to each model.  

\begin{figure}[]
\begin{center}
\plotone{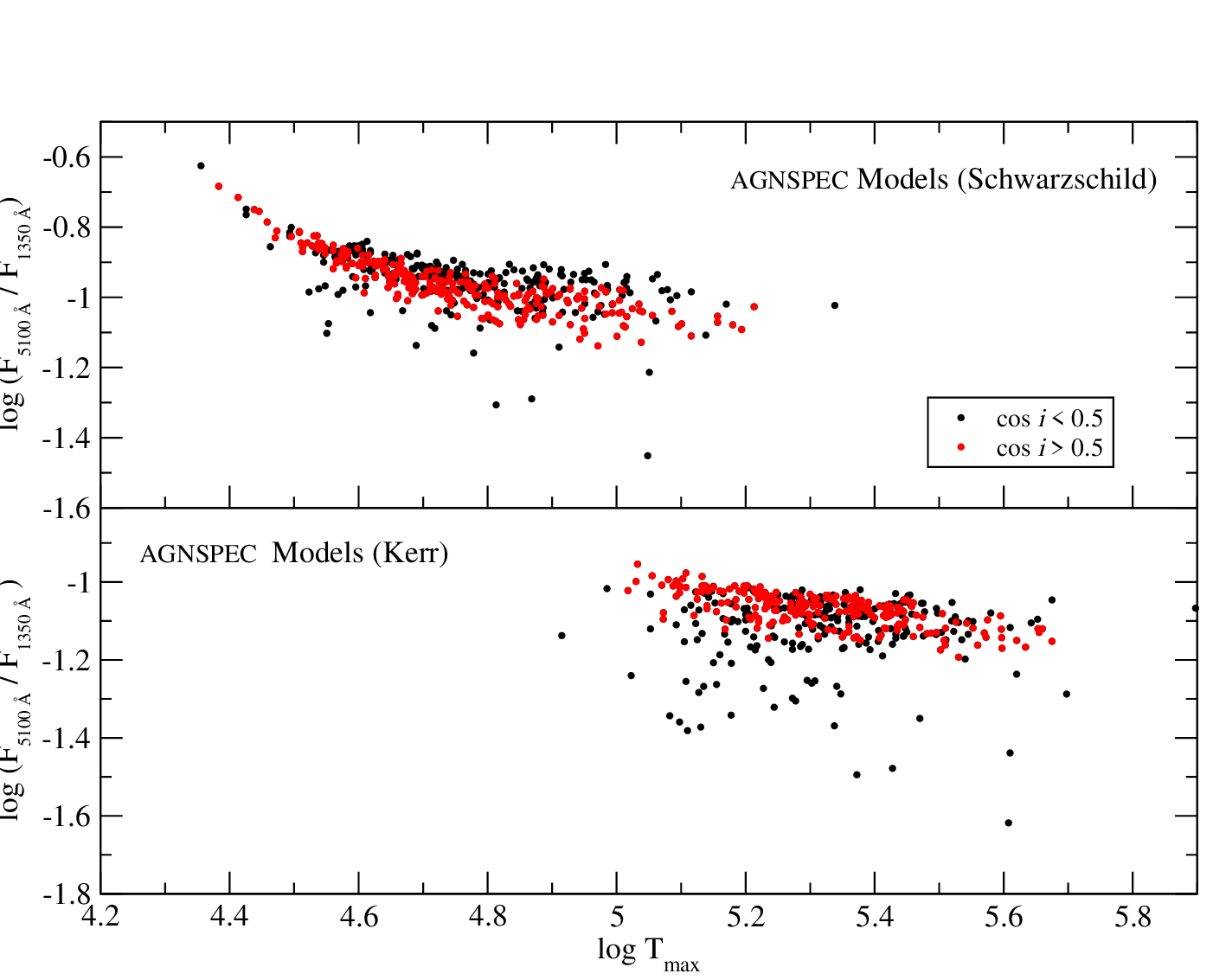} 
\figcaption[fig1]{ Continuum flux ratio
$\flam(\lambda5100)/\flam(\lambda1350)$ for Schwarzschild and Kerr
black holes.  The Schwarzschild models are cooler because they were
computed for the  same set of values of \mbh\ and \lbol.  The points
show an overall trend of bluer colors with increasing  \tmax.  Models
with high inclination ($\cos~i < 0.5$, black dots) show more scatter due to the
relativistic effects on edge-on disks.\label{fig:models} }
\end{center}
\end{figure}

In Fig.~\ref{fig:models} we plot the luminosity ratio 
$L_{\lambda}({5100~\textrm{\AA}})/L_{\lambda}({1350~\textrm{\AA})}$
vs. $T_{\mathrm {max}}$. We distinguish
models with  $\cos\, i > 0.50$.  This is a plausible lower
limit to the inclination angle for  observable quasars if
the ``unified'' model of AGN, involving an obscuring torus, applies to
our QSOs (Antonucci \& Miller 1985). There exists a clear overall trend 
wherein the spectra become more blue with increasing disk
temperature, consistent with intuition. The models with
$\cos\,i < 0.50$ add more scatter, since relativistic beaming and
  Doppler shift effects
result in more flux at shorter wavelengths. However, the overall trend
of bluer color with increasing \tmax\ is still 
evident.  We include models for non-rotating black holes for the same 
\mbh\ and \lbol\ as the Kerr models.  These have lower \tmax\ 
 by 0.46~dex for given \mbh\ and \lbol, but they conform to the same
trend of color with \tmax. 
\subsection{The Color-$T_\mathrm {max}$ Relation from the SDSS Datasets}
\label{sec:data} 
\begin{figure}[]
\begin{center}
\plotone{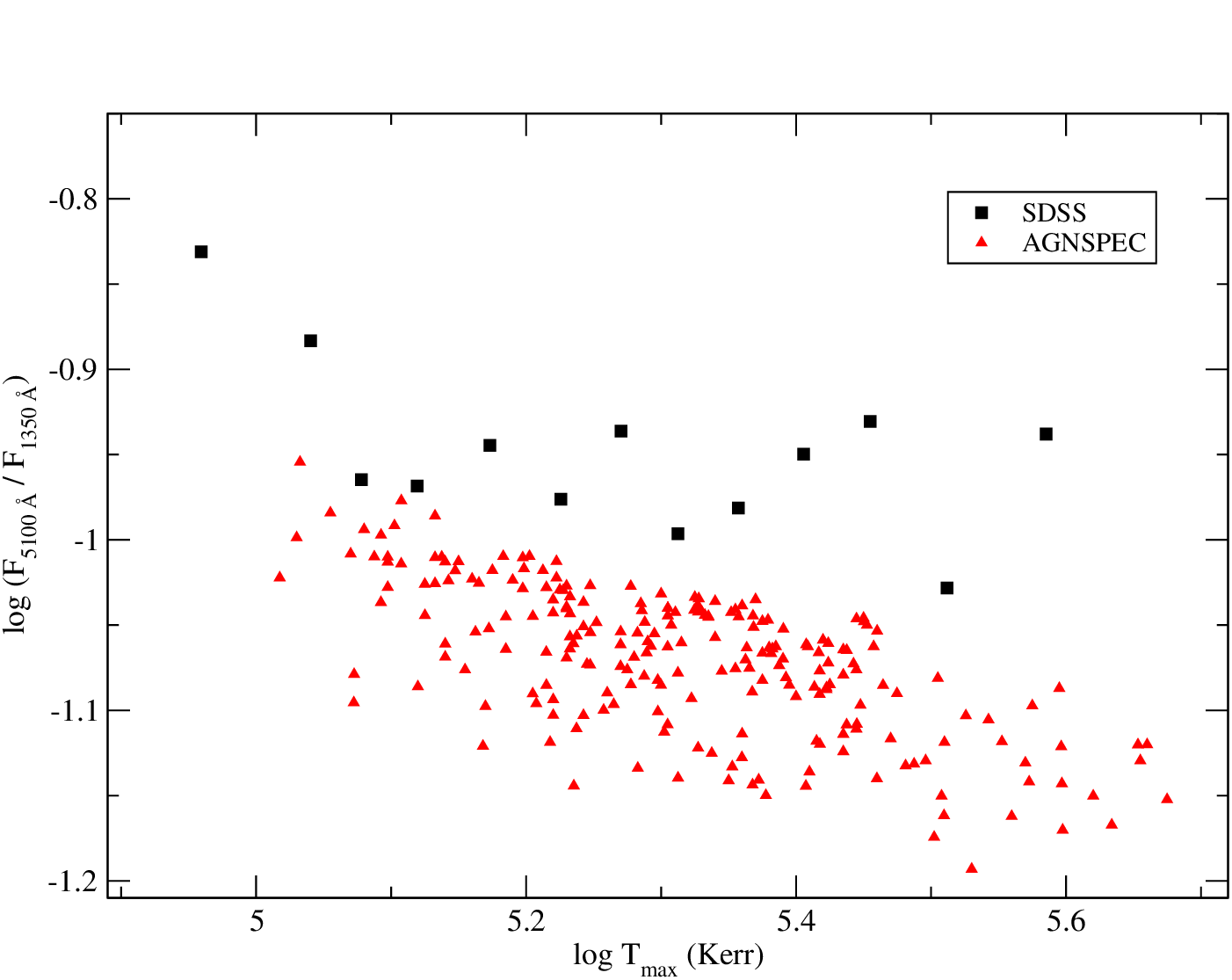}
\figcaption[fig4]{The 5100/1350 \AA\ flux ratios for the $\cos\, i > 0.5$
  Kerr models together with the composite flux ratios for the
  SDSS objects in  \tmax\ bins.   \label{fig:ucurve}
}
\end{center}
\end{figure}

We examined the color (i.e., continuum flux ratios)  of SDSS QSOs
as a function of \tmax\ derived from the FWHM of \mgii\ as in
Eq.~\ref{eq:tmax2} (see Sec.~\ref{sec:tmax}). The SDSS
spectra cover a wavelength range of $\lambda3800$ to $\lambda9200$~\AA.  
The rest-frame colors measured for our sample are $\rlow \equiv \flam(
5100)/\flam (4000)$, $\rmed \equiv \flam (4000)/\flam (2200)$, and
$\rhigh \equiv \flam (2200)/\flam (1350)$.  The objects have redshifts
between $z \sim 0.38$ and  $z \sim 2.0$, such that  \mgii\ lies well
within the SDSS spectrum. The broad \feii\ emission blends
were removed with the aid of templates in the optical and UV bands
(Marziani et al. 2003 and Vestergaard \& Wilkes 2001,
respectively). Continuum fluxes were  computed by taking the mean
(biweight) value of the flux over a 30 \AA\ window centered on the
wavelength of interest.  The chosen wavelengths are relatively
unaffected by \feii\ emission. The FWHM of the broad \mgii\
$\lambda\lambda2796,2803$ doublet was measured by means of a least squares
fit of a Gauss-Hermite function (Pinkney et al. 2003) to the line
profile, together with a linear fit to the continuum in the vicinity 
of the line. Fits were accepted if the line widths and equivalent
widths had an accuracy better than 15\% and visual inspections showed
a good fit and no artifacts in the spectrum or continuum windows.
For more details on the measurement procedure, see Salviander et al. (2006).
\begin{figure}[]
\begin{center}
\plotone{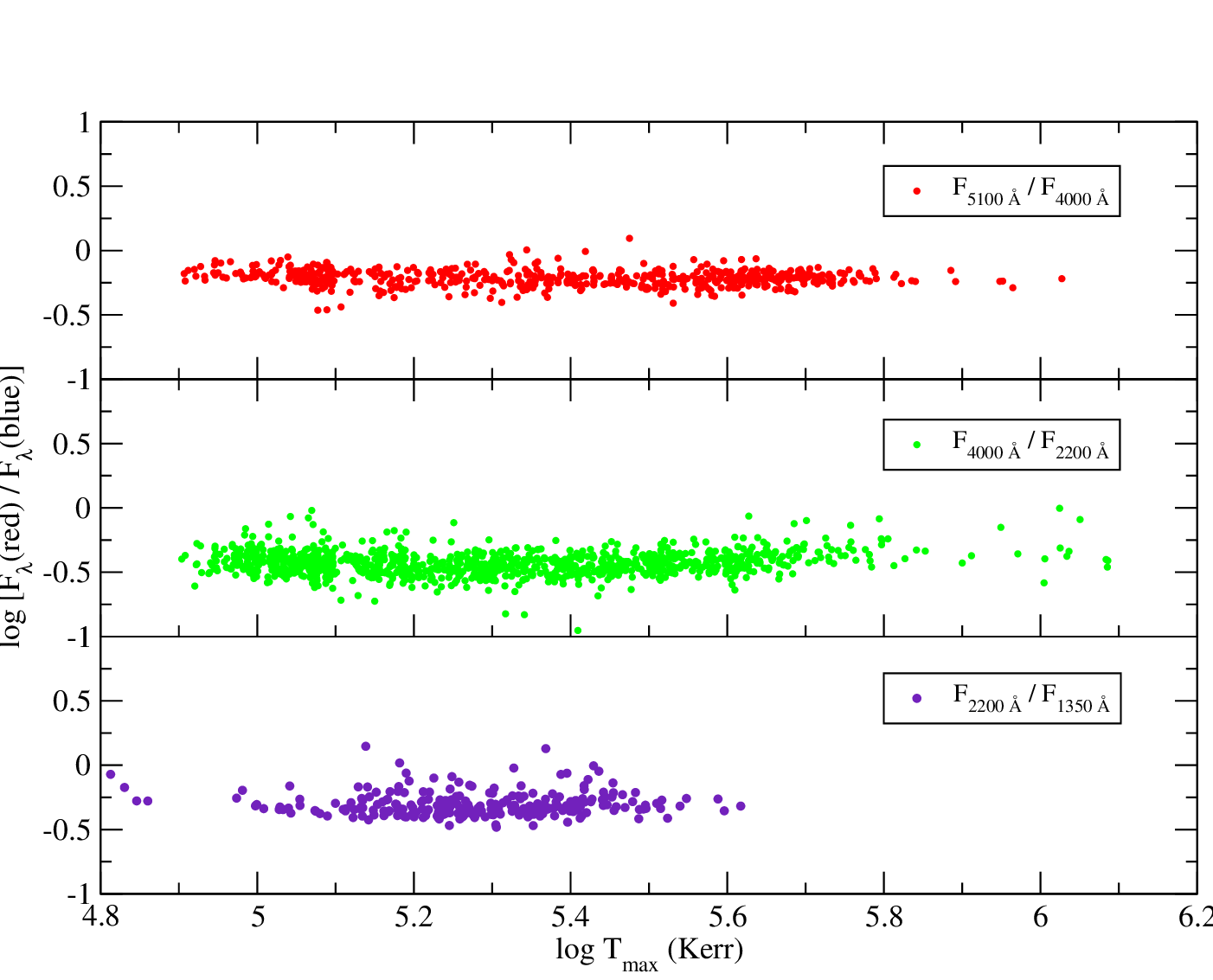}
\figcaption[fig2]{Observed continuum colors for individual QSOs as
  given by flux ratios at three pairs of wavelengths.  The abscissa is log
  \tmax\ derived as described in the text, on the assumption of a
  rapidly rotating black hole. 
\label{fig:alldata}}  
\end{center}
\end{figure}
Figure~\ref{fig:alldata}  shows the results for individual QSOs in the
three wavelength ranges.  The figures show substantial scatter, but
overall a flat or upward trend, particularly for higher \tmax.  In
order to clarify the trends, we have binned the data for each
wavelength range such  that there are approximately equal numbers of
objects per bin of \tmax\ (see Fig.~\ref{fig:breakdown}.  For a given
bin, we averaged the flux 
ratios  $\log\, \rlow$, $\log\, \rmed$, and $\log\, \rhigh$, for each
object for which these data exist.  At longer wavelengths, \rlow\
agrees fairly well with the predictions of the Kerr models. 
The observed \rmed\ tracks the models at low
\tmax\ but shows a significant upturn above $\sim 10^{5.4}$~ K.  Our
shortest wavelength pair, 2200/1350~\AA, disagrees with theoretical
predictions even at low disk temperature, and the departure increases
at higher \tmax~.  The observed trend of \rhigh\ is nearly flat with
\tmax,  not showing the `U'-shape seen in \rmed.
\begin{figure}[]
\begin{center}
\plotone{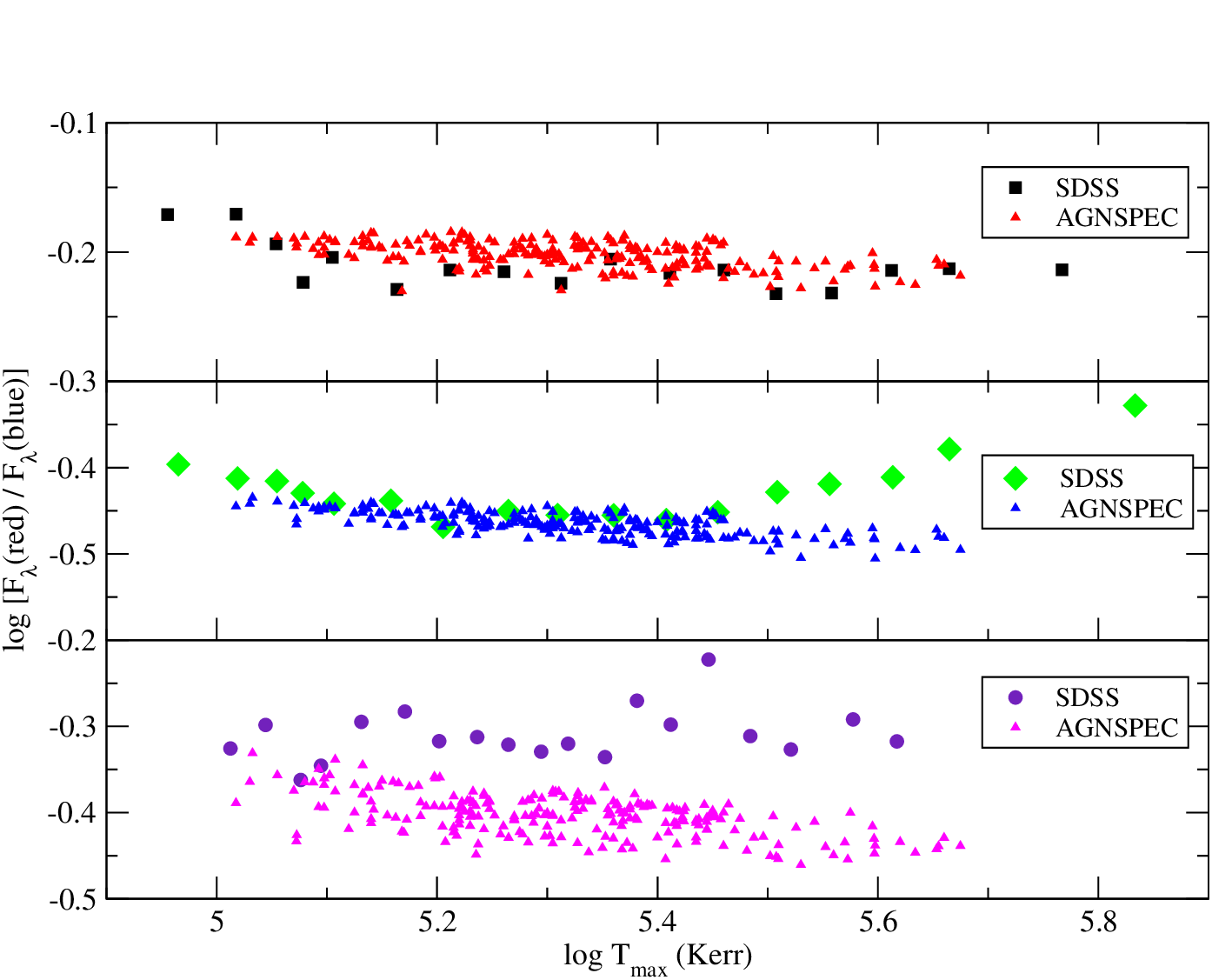}
\figcaption[fig3]{
The three continuum flux ratios (5100\AA/4000\AA, 4000\AA/2200\AA,
2200\AA/1350\AA, as in  Figure~\ref{fig:alldata}) averaged in bins of
\tmax\ and compared 
with results of the disk models (assuming $\cos\, i >
0.5$).\label{fig:breakdown}} 
\end{center}
\end{figure}
In order to display a wider wavelength range than is possible with
SDSS spectra of a given redshift, we have combined the binned
results for the above three pairs of wavelengths by summing the
averages for $\log\, \rlow$, $\log\, \rmed$, $\log\, \rhigh$ to get
a total $\log\, \rall =   \log(\flam(5100) / \flam(1350))$.
Figure~\ref{fig:ucurve} shows the composite color--\tmax\ relation for the
wavelength range 5100~--~1350~\AA\ as well as this luminosity ratio for
individual models calculated as described in
Sec.~\ref{sec:models}. Over this wide wavelength trend, the models
show a substantial trend toward bluer colors with increasing
temperature.  At low \tmax, there is 
rough agreement between the observations and models, whereas at higher
temperatures, the observed colors remain fairly constant and
depart increasingly from the models. Richards et al.
(2003) find a trend of narrower of H$\beta$ with redder continuum
slope, consistent with our findings.  They note the possible
connection with black hole mass (however, they find a
 trend of weaker, not narrower \mgii\ for the same sample).
\section{Discussion}
\label{sec:discussion}
We have found that QSOs show mixed agreement with the predictions of
accretion disk models as a function of disk temperature \tmax.  We
consider the three wavelength pairs $r_{54}$, $r_{42}$, and $r_{21}$
involving ratios of \flam\ at 5100/4000, 4000/2200, and 2200/1350~\AA,
respectively.  Accretion disk models with random inclinations show the
expected trend of bluer colors with increasing \tmax.   Observed
values of $r_{54}$ agree well with the models.  For $r_{42}$, the
observations agree with the models for $\tmax~<~10^{5.4}~K$, but for
higher temperatures the observed colors become redder while the models
become bluer.  For $r_{21}$ at low \tmax, the observed colors are
slightly redder than the models. The observed color remains
roughly constant with increasing \tmax\ while the models become
distinctly bluer.  Thus, for both $r_{42}$ and $r_{21}$, the observed
QSOs fail to show the expected trend with \tmax.  Possible
explanations of this discrepancy might include erroneous values of
\mbh\ and \mdot, host galaxy contamination, dust reddening,  and
inadequacy of the disk models.  
\subsection{Reliability of disk parameters}
\label{sec:param}
The derived values of \tmax\ depend on \mbh\ and \mdot.  Values of
\mbh\ from echo measurements of the BLR radius (Kaspi et al. 2005,
and references therein) agree with host galaxy velocity dispersions
(Gebhardt et al. 2000b; Nelson 2000).   Comparison of black hole
masses from an expression like Eq.~\ref{eq:mbh}  with
host galaxy luminosities  and velocity dispersion \sigstar\ indicates
that \mbh\ is reliable in the mean to  $\sim0.1$ dex with a
dispersion of  $\sim0.4$ dex in individual objects (McLure \& Dunlop
2002; Greene \&  Ho 2005).  The resulting scatter of 0.2~dex and
zero-point  uncertainty of 0.05~dex in \tmax\ does not significantly
affect our  conclusions. 

More difficult to assess is the uncertainty in \mdot.
We used our disk models to assess the error introduced
from using a constant bolometric correction  $f_L = \lbol/\lamLlam$
over our range of \tmax.  The true bolometric luminosity
is given by \mdot\ and $\epsilon$ for the assumed black hole spin.  We
find that the model-derived $f_L$ generally exceeds 9 in a way
dependent on inclination angle. For Kerr holes, taking only inclination 
angles of less than 60$\degr$, we find values of $f_L$ ranging
up to $\sim$80. On average, our assumed $f_L$ of 9 underestimates the
true value of $f_L$ by $\sim$ 0.4 dex. Models with very small
\cosi\ can have values of $f_L$ of up to $\sim$200-300.
Such objects should have a much higher \tmax\ than we have assigned;
but if such QSOs can be observed at all, they will represent only a
small number of objects.  Since the bolometric correction factor
enters into the formula for \tmax\ in the one-fourth power, the effect
of an average 0.4 dex underestimation of the luminosity
translates into a 0.1 dex underestimation of the temperature.

\begin{figure}[]
\begin{center}
\plotone{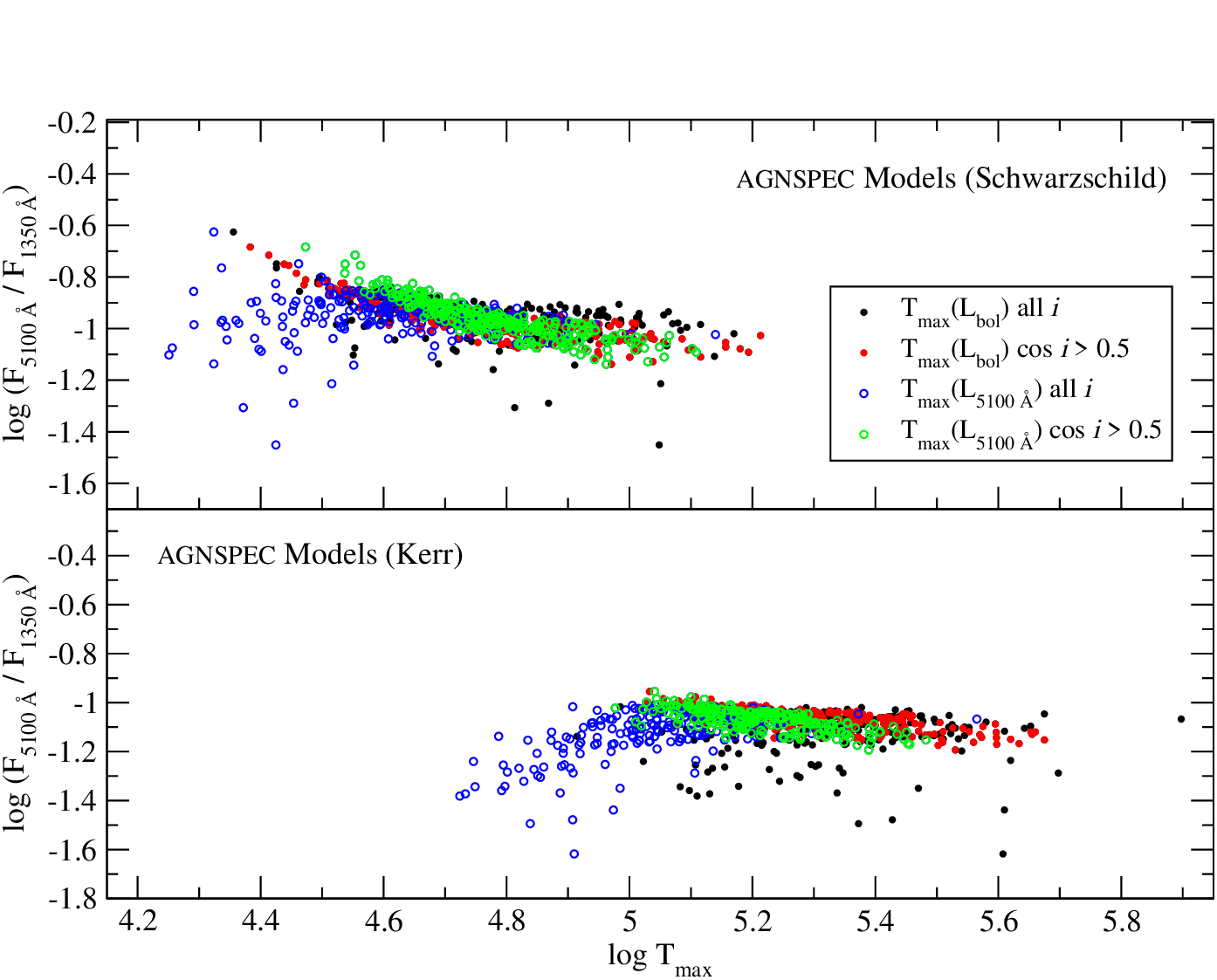}
\figcaption[fig5]
{
The same models as computed for Fig.~\ref{fig:models} but
 with  \tmax\ computed from the model spectrum using $\lbol = 9 \times
 \lamLlam$. Disks with high inclination show
 significant under-estimation of  \tmax\ (see discussion of $f_L$ in text). Low
 inclination disks, however, retain the qualitative
 color v. \tmax\ trend  seen when the model colors are plotted as a
 function of the true  \tmax. \label{fig:mad} }
\end{center}
\end{figure}

Figure~\ref{fig:mad} shows a comparison between the models
of Fig.~\ref{fig:models} and those same models with \tmax\ computed from 
$\lbol = 9 \times \lamLlam$. The effect of underestimating \lbol\ for
objects of low inclination is slight, as noted above. The effect of
significant underestimation of \lbol\ for high inclination objects is
evident. These objects' \tmax\ are far too low compared with their true
temperatures. While these objects do show a trend of increasing red color
with \tmax, this only manifests at low \tmax\ and only in objects of
high inclination.
Our discussion has mostly assumed a near-extremal Kerr
black hole ($a_\ast = 0.998$). It is not known what the black hole
spin distribution is in quasars; the spin parameter is dependent on
the merger history of the black holes (which tends to decrease the
spin) and the gas accretion history (which will spin up the accreting
black hole). It is expected to range from $\sim$0.6 up to about
$a_\ast = 0.9$ (see, e.g. Volonteri et al., 2005; Gammie, Shapiro, \&
McKinney, 2004; Hughes \& Blandford 2003). 
The effect of assuming a variety of spins for our observations
would be to lower \tmax\ by up 0.46~dex (for a zero spin black
hole). This may serve to more closely align the lower temperature bins
with the models, but only increases the discrepancy of the
observations at higher \tmax.  Uncertainties in \tmax\
whether due to uncertainties in \lbol\ or overestimating the black
hole spin do not seem likely to resolve the discrepancy between
observed colors and disk models. 

\subsection{Host galaxy contamination}
\label{sec:host}
Another contributor to QSO colors is starlight from the host galaxy.
The contribution of host-galaxy light is indeed seen via stellar
absorption lines in SDSS composite spectra (Vanden Berk et  
al. 2001).  They find a contribution to the composite quasar
continuum from stars of about 7\%--15\% at the locations of Ca~II
$\lambda$3933 and Na I $\lambda$5896 and about 30\% at the locations
of Ca II  $\lambda\lambda$8498, 8542. Studies of black holes in nearby
quiescent galaxies find \mbh\ roughly proportional to galaxy
luminosity (see review by Kormendy \& Gebhardt 2000).  If we assume
$L_{gal} \propto \mbh$, we expect an increasing galaxy contribution
with decreasing 
$L/\led$.  We show in Fig.~\ref{fig:lled} that for our SDSS QSOs, 
$L/\led$ increases with \tmax, so the galaxy contribution should make
the observed QSOs increasingly redder than the models for decreasing \tmax,
opposite to the trend that we find. Finally, we have made composite
spectra for our QSO sample grouped according to \tmax.   
The Ca K line, visible but weak, suggests a galaxy contribution of only a few
percent at $\lambda4000$, and no perceptible difference between the
objects with higher and lower \tmax.  This suggests that host galaxy
starlight does not cause the divergence of observed and model colors
with increasing \tmax.

\subsection{Reddening}\label{s:red}
Numerous studies have considered the degree of reddening of AGN and
its wavelength dependence (e.g., Wilkes et al. 1999; Richards et
al. 2003).  Hopkins et al. (2004) studied the
reddening in QSOs of SDSS DR1.  Based
on correlations of color and 
spectral curvature and reddening curves by Pei (1992), they conclude
that the minority of objects that do show significant reddening are
consistent with an SMC reddening curve but not with LMC or Galactic
reddening.  However, the patterns in Fig.~\ref{fig:breakdown} are 
not consistent with
reddening by dust.  The observed values of $r_{42}$ follow the Kerr models
below $\tmax = 10^{5.4}$~K, then turn up sharply toward redder values.  In
contrast, $r_{21}$ deviates from the models even at the lowest
temperatures, showing roughly constant values as a function of  \tmax.  
For SMC dust, the reddening of the $r_{54}$ flux ratio is relatively small and
inconclusive for our discussion.  The argument is even stronger for
LMC or MWG dust, for which the reddening between $\lambda1350$ and
$\lambda2200$ is smaller or even negative.   The Gaskell et al. (2004)
reddening curve is flat in the ultraviolet, giving similar
values of differential extinction of $r_{21}$ and $r_{42}$, and much less for
$r_{54}$.  Such a reddening curve cannot explain the results in
Fig.~\ref{fig:breakdown}.   

\begin{figure}[]
\begin{center}
\plotone{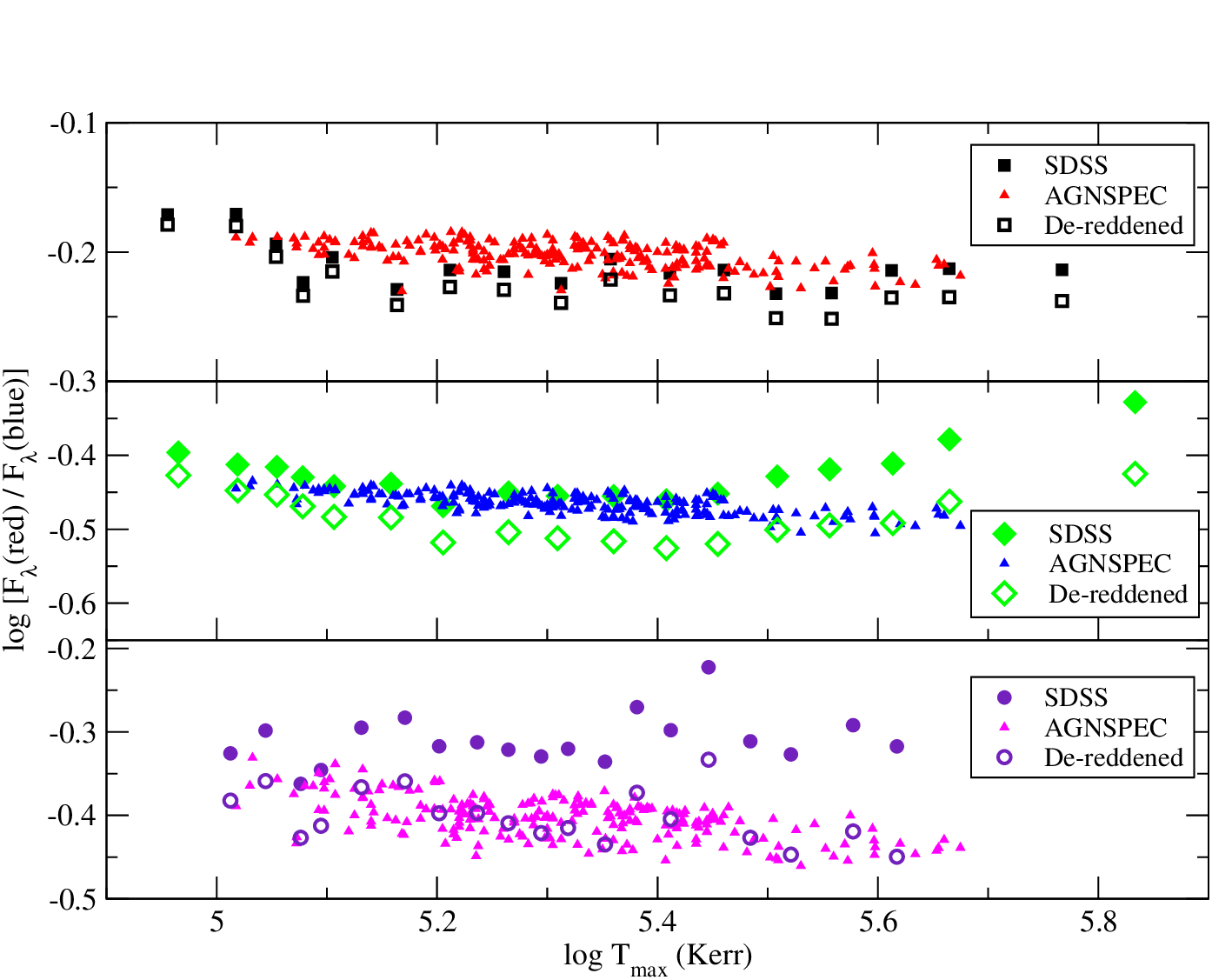}
\figcaption[fig6]{
The binned observed points of
Fig.~\ref{fig:breakdown} de-reddened
with an SMC extinction curve. We assume here that the difference in 
the 2200~\AA/1350~\AA~flux ratio between the observed points and the
models is accounted for by  reddening correlated with broad line-width.
The other two flux ratios are then de-reddened by the corresponding amount
according to the SMC extinction curve (see text).
\label{fig:red} }
\end{center}
\end{figure}

We caution that the amounts of reddening required to affect the curves
in Fig.~\ref{fig:breakdown} is not large.  An increase in $r_{21}$ of
$\sim0.09$~dex requires $E(B-V) = 0.03$ or $A_V = 0.09$;  the same
change in $r_{42}$ requires $E(B-V) = 0.05$ or $A_V = 0.16$.  For
comparison, the amount of Galactic reddening of the SDSS QSOs used
here averages about $A_g = 0.13$ for the SDSS $g$-band photometric
color.   However, the reddening would need to correlate systematically
with broad line width to explain the trends in $r_{42}$ or
$r_{21}$. To test this possibility, we assume that the
amount of dust   reddening correlates with broad line width
(i.e. \tmax) in such a  way as to give overall agreement in the
ratio $r_{21}$.  
Figure~\ref{fig:red} shows the binned points of
Fig.~\ref{fig:breakdown} de-reddened  with an assumed reddening of
$r_{21}$ given by $\Delta\,\log\,r_{21} =  -0.57 + 0.125 \cdot
\log{\tmax}$.  This is 
tailored in such a  way as to give overall agreement between the
de-reddened observations and the models for $r_{21}$.   The SMC
extinction curve (Pei 1992) gives
$\Delta\,\log\,r_{42}/\Delta\,\log\,r_{21} = 0.61$ and  
$\Delta\,\log\,r_{54}/\Delta\,\log\,r_{21} =  0.16$.  The figure shows
that the de-reddened 
$r_{42}$ now is bluer than the models for lower \tmax, and at high
\tmax\ it rises to redder  values in a way not reflected in the
models. The de-reddened $r_{54}$ now is significantly bluer than the
models for most values of \tmax.   Evidently, reddening cannot
reconcile the models and observation in detail, even for a carefully
tailored increase of reddening with broad line width.  

Reddening may nevertheless have a significant effect on the observed
colors and their relationship to the model colors. For example,
Constantin \& Shields (2005) find evidence for  
greater reddening in Narrow Line Seyfert 1s (NLS1s) than in the AGN
population as a whole. Baskin \& Laor (2005) plot the \civ/\hbeta\  
flux ratio against continuum slope and find evidence for  reddening
amounting to several tenths dex in the line ratio.  Even
if it does not correlate significantly with \tmax, reddening of this
magnitude would significantly alter the relationship of the observed
colors and disk models, as in our Fig.~\ref{fig:breakdown}.
Willott (2005) argues that the reddening curve proposed by Gaskell et
al. is affected by selection biases relating to the composite spectra
used.  Higher redshift objects define the short wavelength part of the
composite,  and 
there may be a selection for lower extinction at higher redshift
because of the limiting magnitude of the survey providing objects for
the composite.  Such a bias might affect our results, which use higher
redshift objects for $r_{21}$.  However, this bias would cause the
extinction to be lower for $r_{21}$ than $r_{42}$, opposite to the
result in Fig.~\ref{fig:breakdown}.

Baker (1997) finds that the radio-loud quasars (RLQs) are redder for
more edge-on 
inclinations. Wills \& Browne (1986) find broader emission lines in
RLQs seen more edge-on, as would occur if
\hbeta\ lines were emitted from a disk-like structure.  This would 
result in broader lines for more edge-on disks, and greater reddening
for broader lines.  This is opposite to the trends in Fig.~\ref{fig:breakdown}.
While most SDSS QSOs are radio quiet, the same reasoning should apply. 
The relationship between nuclear gas supply and  luminosity is
uncertain. Plausibly, more abundant  gas might lead to greater
reddening as well as greater fueling and higher values of \lbol\ and \tmax.

\subsection{Disk physics}\label{s:physics}
\begin{figure}[]
\begin{center}
\plotone{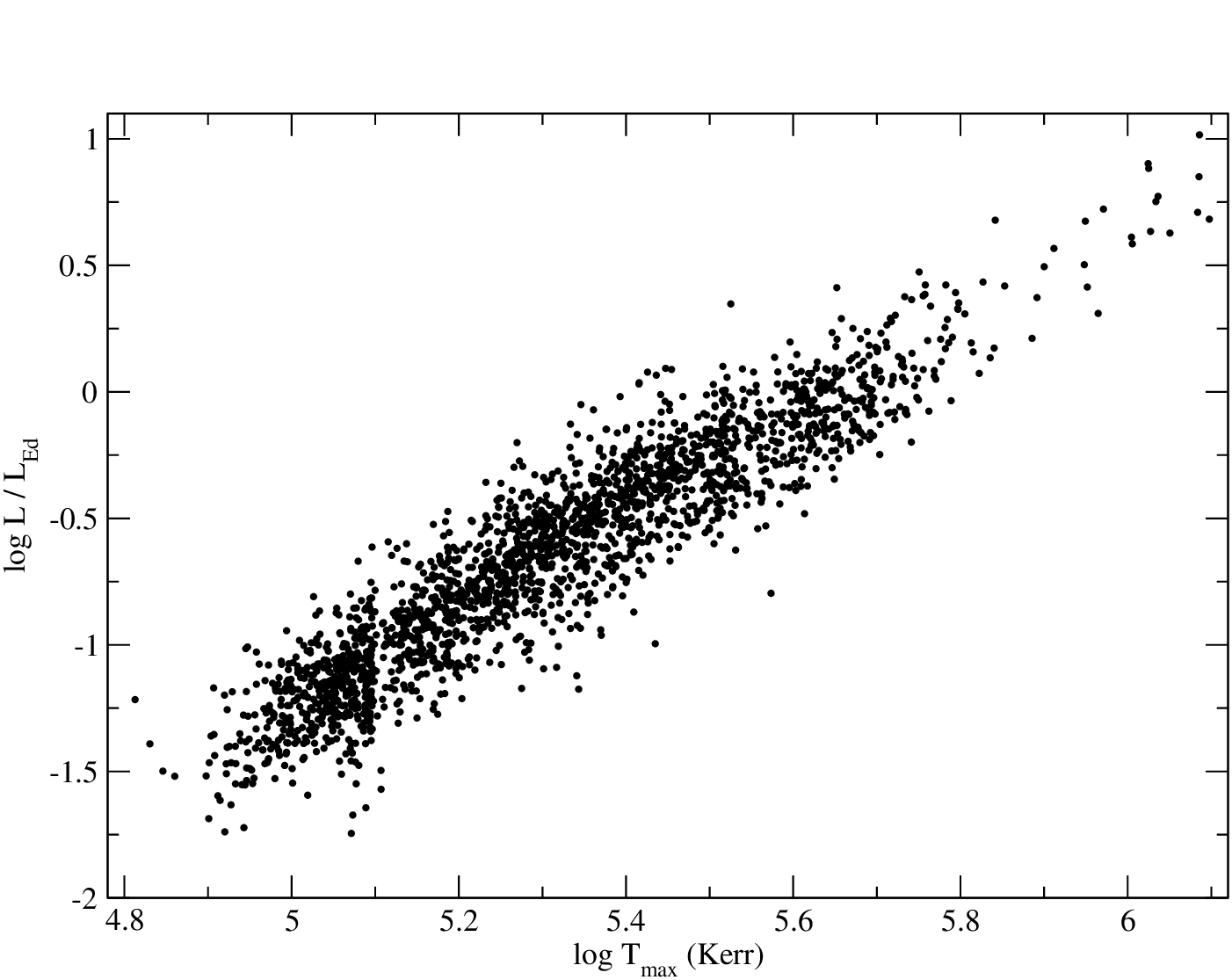}
\figcaption[fig7]{
The Eddington ratio (in log). For higher redshift objects
where $\lambda5100$ is inaccessible, we scale the 
4000~\AA\  continuum to 5100~\AA\ as a power law  with $\alpha_\nu =
-0.44\pm0.1$ (Vanden Berk et al. 2001). The uncertainty has only a
$\sim 1\%$ affect on the derived \mbh\ and \lbol/\led .  
\label{fig:lled} }
\end{center}
\end{figure}

Does accretion disk physics underly the difference between observed colors
and   the models?  Figure~\ref{fig:lled} shows  the strong correlation
of \tmax\ and $\lbol/\led$ for our sample, largely reflecting
  the fact that lower \tmax\ objects have broader lines, and hence
  larger black holes.
(Note that the derived $\lbol/\led$ is
proportional to the bolometric correction $f_L$ that we have assumed). 
Collin \& Hur\'e (2001) and Collin et al. (2002) have discussed high
Eddington  ratios in AGN based on objects with measured reverberation
masses for the black hole.  They note the possible role of slim disks
or non-standard disks in super-Eddington accretion
scenarios.  The largest accretion rates correspond to the narrowest
broad lines, particularly in NLS1
objects. In Fig.~\ref{fig:breakdown}, the common NLS1 definition
FWHM~$\leq$~2000~\kms\ corresponds to log \tmax$\gsim$5.6.

Comparing Fig.~\ref{fig:lled} with Fig.~\ref{fig:breakdown},  we
see that the deviation of $r_{42}$ from the models occurs for
$\lbol/\led > 0.3$ absent significant reddening (see above). This is
just the range where the inner disk is severely thickened by 
radiation pressure, and departures from a thin disk geometry and
atmospheric structure might be expected.  The concept of  ``slim''
accretion disks (Abramowicz et al. 1988) may therefore be relevant.
These models modify the standard equations for thin accretion disks
(Shakura \& Sunyaev 1973) by including terms allowing for the radial
pressure gradient and radial advection of heat.  Szuszkiewicz, Malkan,
\& Abramowicz (1996) discuss the application of slim disks to AGN
energy distributions, including the soft X-ray excess often observed
in AGN (e.g., Turner \& Pounds 1989).  Slim disks are thin at large
radius, and then  increasingly depart from the thin disk solution at
smaller radii as the local radiation flux counteracts the vertical
gravity and thickens the disk.    Slim disks thicken at a transition
radius $r_{\rm slim}$ that increases with $\lbol/\led$.  Thus, longer
wavelength radiation, emitted from the cooler, outer disk may obey the
thin disk solution, while shorter wavelengths originating at smaller
radii depart from the thin disk prediction.  We attempted to estimate
the effect of a slim accretion disk by applying an inner cutoff to the
disk at the radius where the ratio of disk height to radius is $H/R
\approx 0.5$ (see discussion of $H/R$ in Hubeny et al. 2000 and Laor
\& Netzer 1989).  This is based on the idea that the locally radiated
flux and effective temperature may be depressed at radii where
advection carries energy inward.  The \llam(5100)/\llam(4000) ratios
are relatively unaffected by an inner disk cutoff;  however, the color
ratios, $r_{42}$ and $r_{21}$ do show an upward trend at higher
\tmax. Although they do not reproduce the correct magnitude of the
ratios,  $r_{42}$ and  $r_{21}$ show a flat or `U'-shaped curve. The
result of this simple approximation suggests that an accurate
simulation of slim accretion disks could illuminate our observations
of objects with high Eddington ratio.

Comptonization can affect the observed spectra of AGN, involving
either an hot corona overlying the disk or electron scattering
in the atmosphere itself (Shang et al. 2005; Hubeny et al. 2001: and
references therein).
The disk models here do not include Comptonization.  Blaes
et al. (2001), in a study of accretion disk models of 3C 273,
find that Comptonization in the atmosphere does not affect
the optical and mid-UV spectrum.  Nevertheless, Comptonization
should be kept in mind in studies of the QSO energy distribution.

We have considered only the optical and ultraviolet energy distribution.
The statistical sampling allowed by SDSS makes it possible to
study fairly subtle trends in the colors with \tmax.  Nevertheless,
\tmax\ has a much larger effect on the flux at short wavelengths.
Laor et al. (1997) discuss the soft x-ray properties of AGN.
They find that accretion disk models, for reasonable distributions of
\mbh\ and \mdot, predict a larger range of optical to soft x-ray slope
$\alpha_{ox}$ than is observed.  Our Figures~\ref{fig:breakdown} and
\ref{fig:ucurve} likewise show a smaller range of color in the
observations than in the models.

Boroson and Green (1992) identified a number of correlations between
observable properties of QSOs, the strongest of which they designated
``Eigenvector~1".   This includes a tendency for stronger permitted
\feii\ broad line emission together with weaker narrow \oiii\  lines.
Boroson and Green suggested that Eigenvector 1 was driven by Eddington
ratio, in the sense of increasing \feii\ with increasing $\lbol/\led$.
This relationship was confirmed by Boroson (2002), but the physical
causes remain unclear.  The correlation of $\lbol/\led$ 
with \tmax\ suggests that \tmax\ may play a role in the physics
underlying Eigenvector~1.

\acknowledgements
We are greatly indebted to Ivan Hubeny for making available the
{\sc agnspec}  code, which includes the program {\sc kerrtrans} of Eric Agol to
calculate the  relativistic transfer function. 
We thank Shane Davis, Robert Antonucci, Omer Blaes, Mike Brotherton,
Ivan Hubeny, Pamela Jean, Zhaohui Shang, and Bev Wills for
helpful discussions and assistance.   
EWB is supported by Marie Curie Incoming European Fellowship contract
MIF1-CT-2005-008762 within the 6th European Community Framework Programme.
This research was supported in part by the National Science Foundation
under Grants No. PHY99-07949  and AST-0098594, and
in part by the Texas Advanced Research Program under grant 003658-0177-2001.

Funding for the Sloan Digital Sky Survey (SDSS) has been provided by
the Alfred P. Sloan Foundation, the Participating Institutions, the
National Aeronautics and Space Administration, the National Science
Foundation, the U.S. Department of Energy, the Japanese
Monbukagakusho, and the Max Planck Society. The SDSS Web site is
http://www.sdss.org/. The SDSS is managed by the Astrophysical
Research Consortium (ARC) for the Participating Institutions. The
Participating Institutions are The University of Chicago, Fermilab,
the Institute for Advanced Study, the Japan Participation Group, The
Johns Hopkins University, the Korean Scientist Group, Los Alamos
National Laboratory, the Max-Planck-Institute for Astronomy (MPIA),
the Max-Planck-Institute for Astrophysics (MPA), New Mexico State
University, University of Pittsburgh, University of Portsmouth,
Princeton University, the United States Naval Observatory, and the
University of Washington.

\end{document}